\documentclass{PoS}

\newcommand\text[1]{\mathord{\hbox{#1}}}

\newcommand\doi[1]{\href{http://dx.doi.org/#1}{doi:#1}}

\newcommand\url[1]{url:\href{#1}{#1}}

\title{The QCD Equation of State}

\ShortTitle{The QCD Equation of State}

\author{\speaker{Tanmoy Bhattacharya}%
         \thanks{for the HotQCD collaboration: A. Bazavov,
           T. Bhattacharya, C. DeTar, H.-T. Ding, S. Gottlieb,
           R. Gupta, P. Hegde, U.M. Heller, F. Karsch, E. Laermann,
           L. Levkova, S. Mukherjee, P. Petreczky, C. Schmidt,
           C. Schroeder, R.A. Soltz, W. Soeldner, R. Sugar, M. Wagner,
           P. Vranas. The speaker was supported by DOE grant No.
           DE-KA-1401020}\\
        Los Alamos National Laboratory\\
        E-mail: \email{tanmoy@lanl.gov}}

      \abstract{Results for the equation of state in 2+1 flavor QCD at
        zero net baryon density using the Highly Improved Staggered
        Quark (HISQ) action by the HotQCD collaboration are
        presented. The strange quark mass was tuned to its physical
        value and the light (up/down) quark masses fixed to \(m_l 
        = 0.05m_s\) corresponding to a pion mass of 160 MeV in the
        continuum limit. Lattices with temporal extent \(N_t=6\), 8,
        10 and 12 were used. Since the cutoff effects for \(N_t>6\)
        were observed to be small, reliable continuum extrapolations
        of the lattice data for the phenomenologically interesting
        temperatures range \(130 \mathord{\rm MeV} < T < 400
        \mathord{\rm MeV}\) could be performed.  We discuss
        statistical and systematic errors and compare our results with
        other published works.}

\FullConference{The 32nd International Symposium on Lattice Field Theory\\
		 23-28 June, 2014\\
		 Columbia University New York, NY}

\begin{document}

\section{Introduction}

Hadronic matter deconfines at temperatures above \(155~\text{MeV}\),
where quark and gluons constitute the relevant degrees of freedon.
Nevertheless, the theory remains non-perturbative because of the
strongly interacting infrared sector, and the equation of state (EOS)
of this quark-gluon plasma requires non-perturbative analysis using
methods such as lattice QCD. The EOS is needed in modeling the
hydrodynamic evolution of the quark-gluon plasma probed in
relativistic heavy-ion collisions and in understanding the cooling of
the early universe.  Phenomenologically, the most interesting region
for lattice QCD analysis is between $T \ sim 140 $~MeV and
\(400~\text{MeV}\), {\it i.e.}, above the validity of hadron-resonance
gas models and spanning the temperatures probed in relativistic
heavy-ion collisions including the `transition' region around
\(154(9)~\text{MeV}\)~\cite{Bazavov:2011nk}.

\section{Lattice Details and the Trace Anomaly}

\begin{figure}[b]
  \begin{center}
    \begin{tabular}{cc}
    \colorbox{white}{\includegraphics[width=0.38\textwidth]{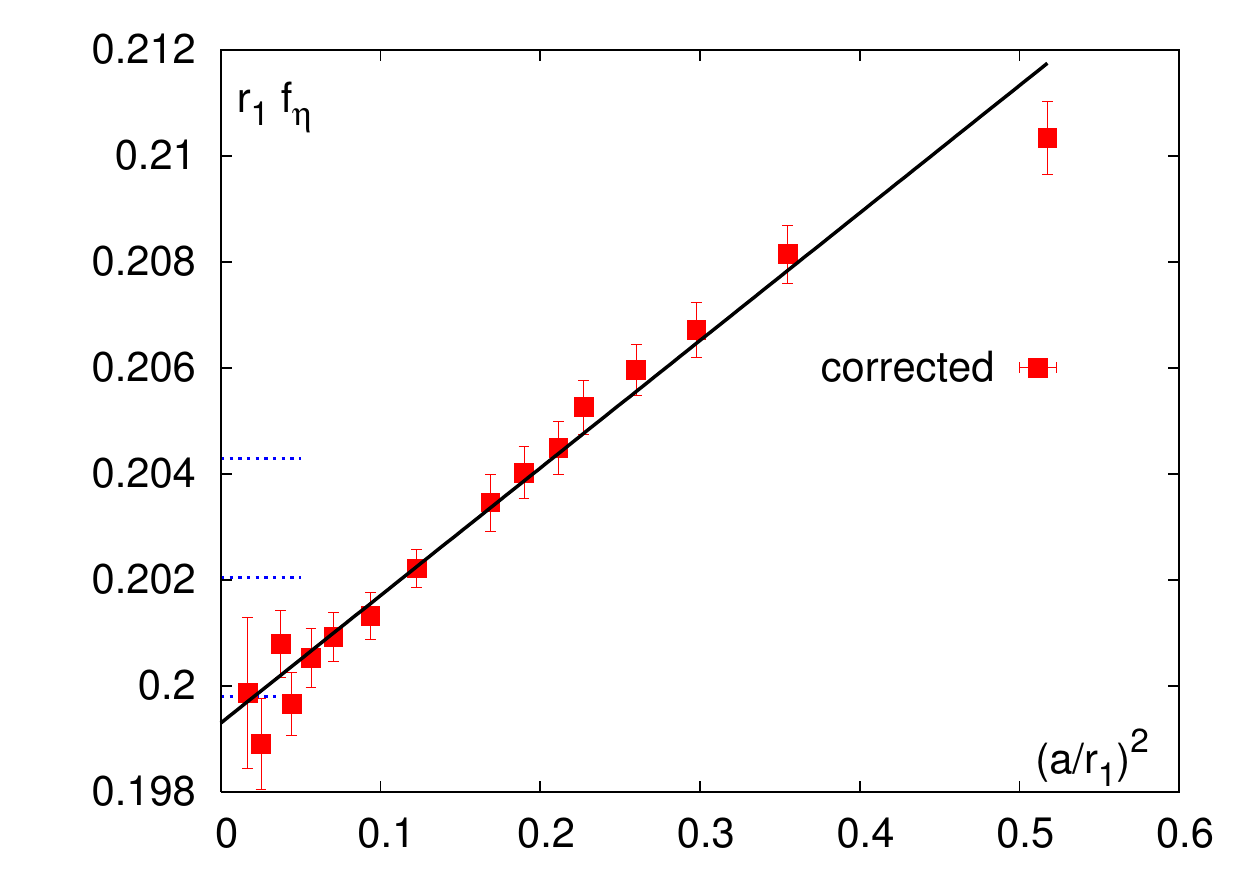}}&
    \colorbox{white}{\includegraphics[width=0.38\textwidth]{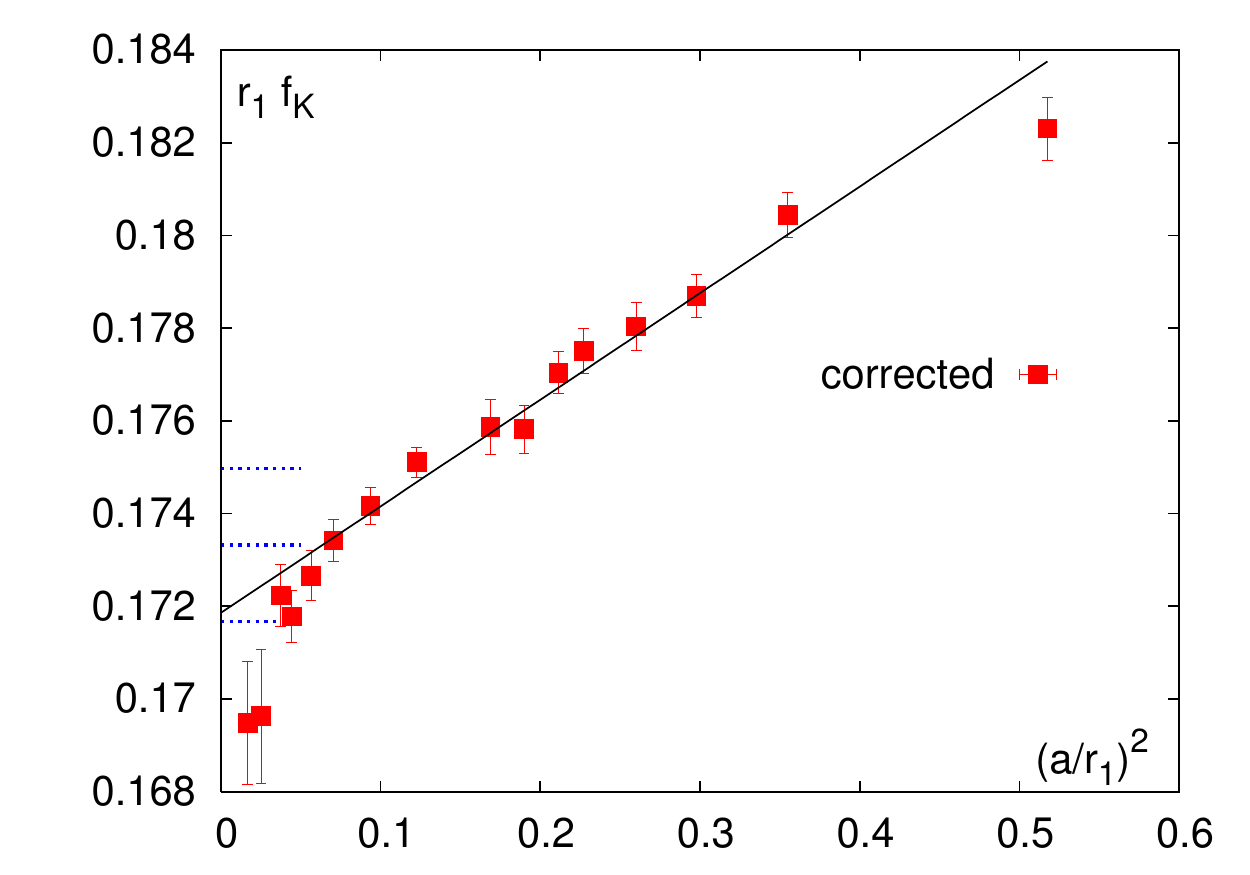}}\\
    (a)&(b)\\
    \colorbox{white}{\includegraphics[width=0.38\textwidth]{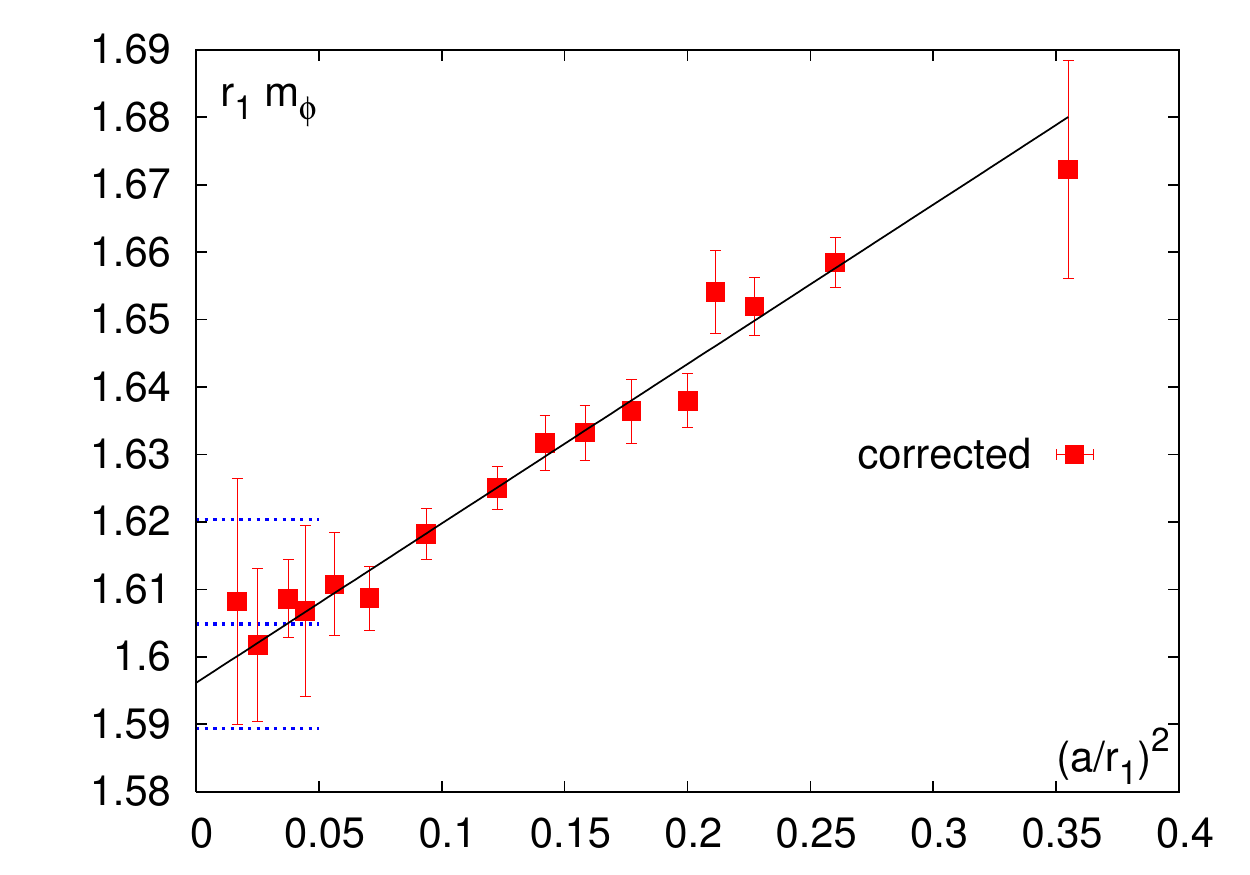}}&
    \colorbox{white}{\includegraphics[width=0.38\textwidth]{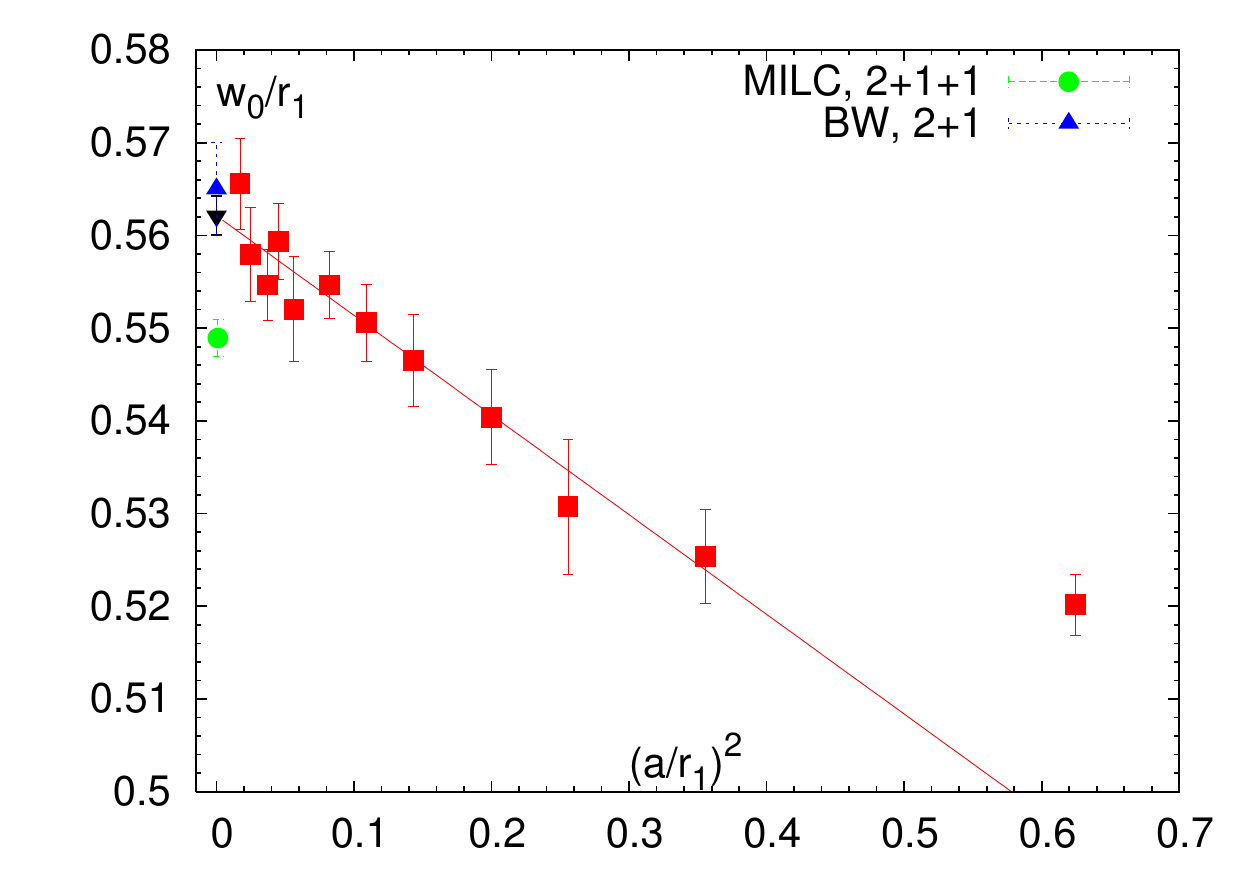}}\\
    (c)&(d)
    \end{tabular}
  \end{center}
  \caption{The continuum estrapolation of (a) \(r_1f_\eta\), (b)
    \(r_1f_K \), (c) \(r_1M_\phi\) and (d) \(w_0/r_1\), along with
  their continuum values~\cite{Davies:2009tsa,Beringer:1900zz}.  The
  data have been corrected for the mass mistuning described later in
  the text.}
  \label{fig:scale}
\end{figure}

We summarize the calculation of EOS~\cite{Bazavov:2014pvz} by the
HotQCD collaboration using \(2+1\) flavors of fermions with the
HISQ/tree action.  In this calculation, we varied \(\beta=10/g^2\) in
the interval \(5.9\)--\(7.825\) along a line of constant physics
(LoCP) defined by the $s\overline s$ meson mass tuned to
\(M_{s\overline s} \approx 695~\text{MeV}\) and the light quark mass
fixed at \(m_l = m_s/20\) (\(M_\pi\approx 160~\text{MeV}\)), as used
in our previous calculation~\cite{Bazavov:2011nk}.  To control
discretization effects, the calculation was done at four values of the
temporal size \(N_\tau = 6\), \(8\), \(10\), and \(12\). The spatial
size was fixed at 4 times \(N_\tau\), {\it i.e.,} \(N_\sigma/N_\tau =
4\).  The lattice update was done with the RHMC algorithm with mass
preconditioning~\cite{Clark:2005sq}.  All the results are reported
with the lattice scale set by \(r_1 =
0.3106(14)(8)(4)~\text{fm}\)~\cite{Bazavov:2010hj}.  We checked that
in our calculation this corresponded to \(r_0=0.4688(41)~\text{fm}\)
and \(w_0=0.1749(14)~\text{fm}\) consistent with the previous
determinations \(0.48(1)(1)~\text{fm}\)~\cite{Aoki:2009sc} and
\(0.1755(18)(4)~\text{fm}\)~\cite{Borsanyi:2012zs} respectively. This
estimate is also consistent, in the continuum limit, with the scale
obtained from various fermionic quantities as shown in
Fig.~\ref{fig:scale}.

\begin{figure}[t]
  \begin{center}
    \begin{tabular}{cc}
      \colorbox{white}{\includegraphics[width=0.38\textwidth]{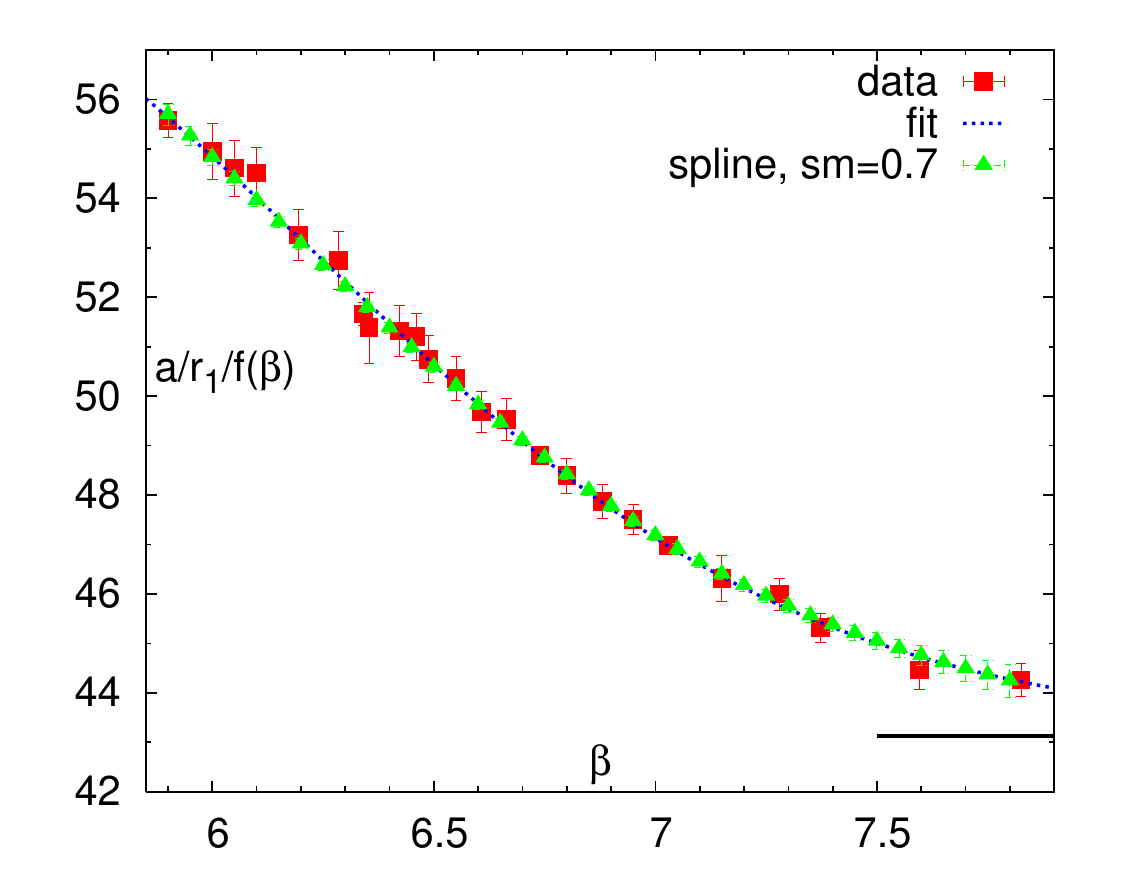}}&
      \colorbox{white}{\includegraphics[width=0.38\textwidth]{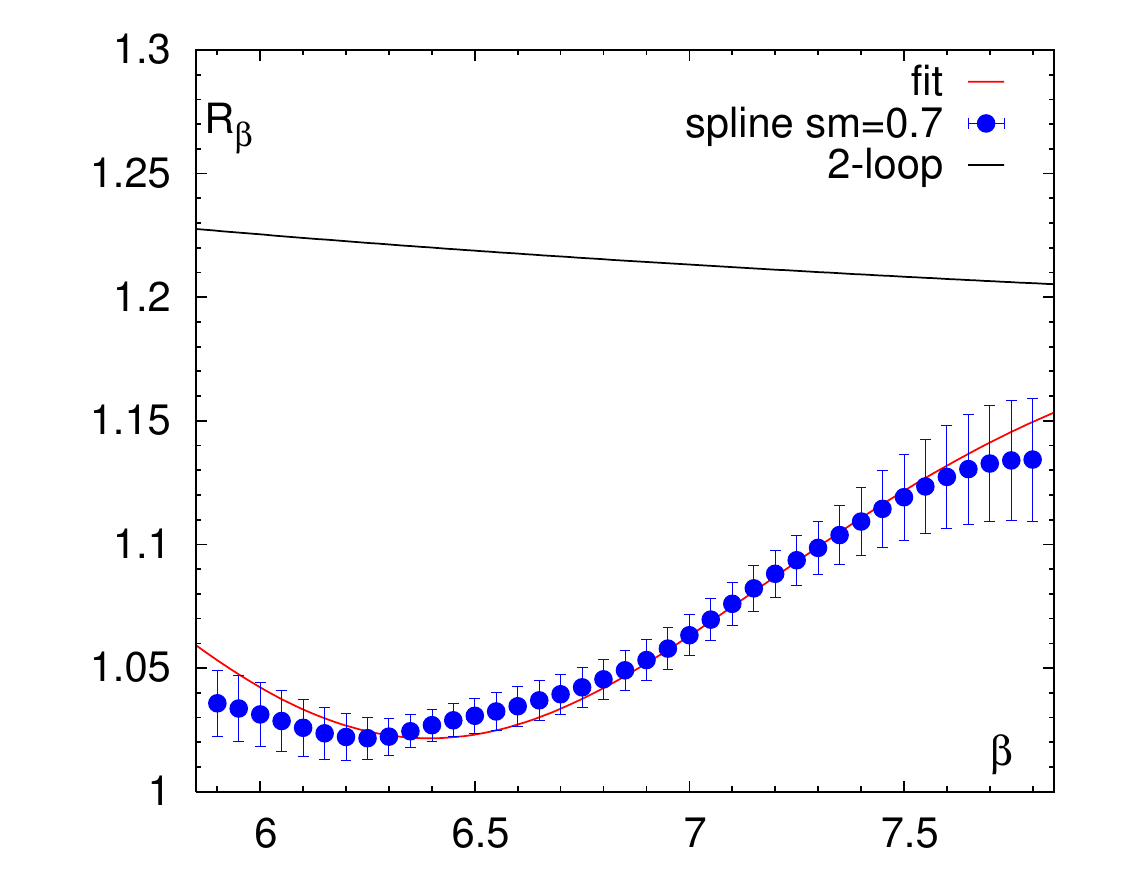}}\\
      (a)&(b)
    \end{tabular}
  \end{center}
  \caption{(a) The scale, \(a/r_1\), normalized by the two-loop beta
    function, \(f(\beta)\), showing a fit and a spline interpolation
    to the data and (b) the non-perturbative $\beta$-function,
    \(R_\beta\), compared to the two-loop value.}
  \label{fig:Rbeta}
\end{figure}

\begin{figure}[t]
  \begin{center}
    \begin{tabular}{cc}
      \colorbox{white}{\includegraphics[width=0.38\textwidth]{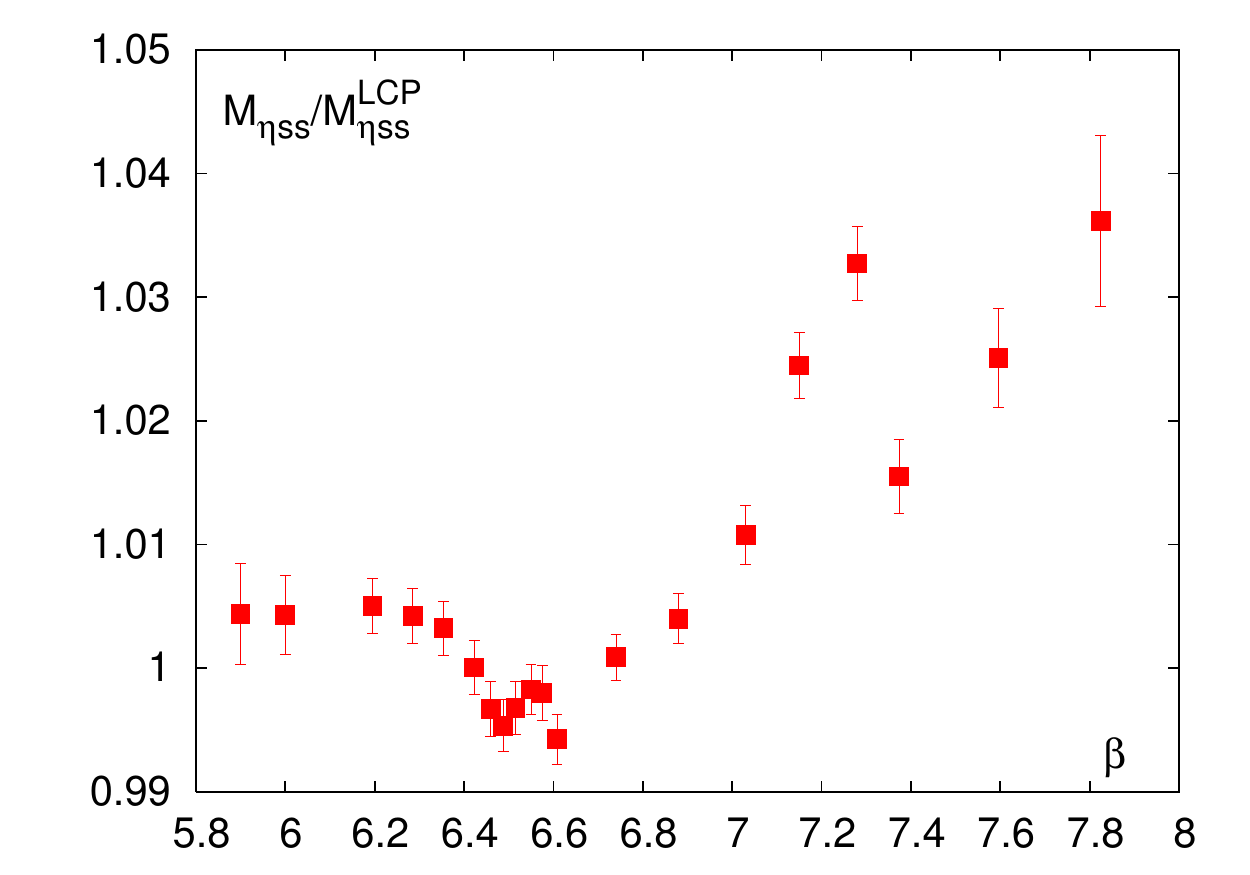}}&
      \colorbox{white}{\includegraphics[width=0.38\textwidth]{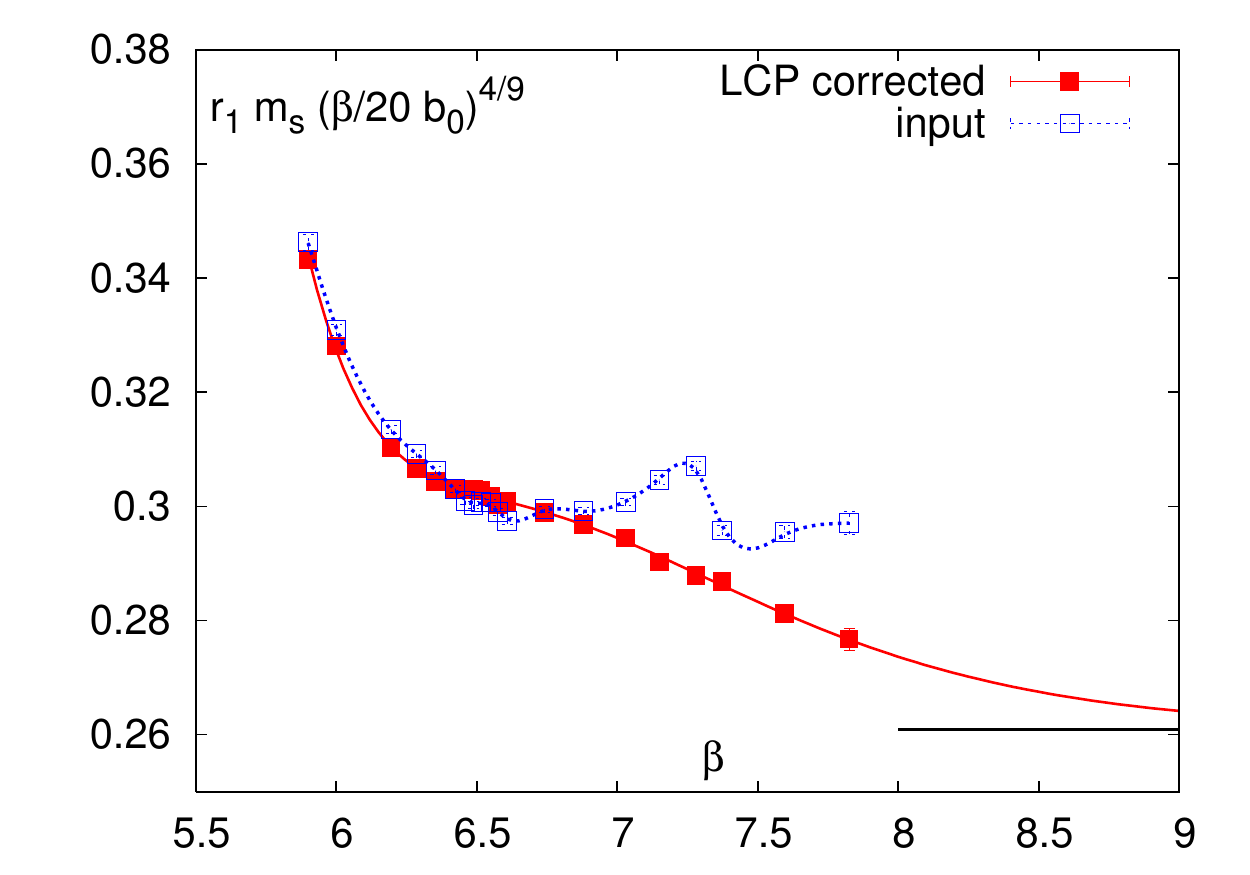}}\\ (a)&(b)
    \end{tabular}
  \end{center}
  \caption{(a) The measured value of \(M_\eta^{ss}\) normalized by
    \(695~\text{MeV}\) showing a few percent mistuning at higher
    \(\beta\). (b) The one-loop renormalization-group-invariant
    strange-quark mass in units of \(r_1^{-1}\), and the same after
    correcting for the mistuning, along with a spline interpolation
    and the asymptotic value at infinite \(\beta\).}
  \label{fig:Mss}
\end{figure}

After integrating out the fermion degrees of freedom, the QCD
partition function on an \(N_\sigma^3N_\tau\) hypercubic lattice can
be written as
\begin{equation}
 Z(\beta,N_\sigma,N_\tau) = \int \prod_{x,\mu} dU_{x,\mu}
 e^{-(\beta\, S_G (U) - S_F(U))}\,,
\end{equation}
where \(s_G(U)\) and \(s_F(U)\) are the gauge and fermionic actions in
terms of the \(\text{SU(3)}\) link variables \(U\).  From this we can
calculate the energy density, \(\varepsilon\), and pressure, \(p\), in
terms of the temperature, \(T\), starting with the trace anomaly,
\(\Theta^{\mu\mu}\) and using the relations
\begin{eqnarray}
  T\frac{d}{d T} \left(\frac{p}{T^4}\right) &=&\frac{\varepsilon -3p}{T^4} \;=\;
     \frac{\Theta^{\mu\mu}_G(T)}{T^4} + \frac{\Theta^{\mu\mu}_F(T)}{T^4}  \,, \\
  \frac{\Theta^{\mu\mu}_G(T)}{T^4} &=& R_\beta \left[ \langle s_G \rangle_0 -
     \langle s_G \rangle_\tau \right] N_\tau^4 \,, \\
  \frac{\Theta^{\mu\mu}_F(T)}{T^4} &=& - R_\beta R_{m}  \left[ 2 m_l\left(
    \langle\bar{\psi}\psi \rangle_{l,0} -
    \langle\bar{\psi}\psi \rangle_{l,\tau}\right)
    + m_s \left(\langle\bar{\psi}\psi
    \rangle_{s,0} - \langle\bar{\psi}\psi \rangle_{s,\tau} \right )
    \right] N_\tau^4 \,,
\end{eqnarray}
where the subscripts \(\tau\) and \(0\) refer to expectation values at
finite and zero temperature respectively, and \(l\) and \(s\) refer to
the light and strange quark condensates, respectively. The function
\(R_\beta\) is the non-perturbative \(\beta\) function,
\begin{equation}
R_{\beta}(\beta) = \frac{r_1}{a} \left( \frac{{\rm d} (r_1/a)}{{\rm d}
  \beta} \right)^{-1}
\end{equation}
shown in Fig.~\ref{fig:Rbeta}
and \(R_m\) is the mass renormalization function,
\begin{equation}
  R_m(\beta) = \frac{1}{m_s(\beta)} \frac{{\rm d} m_s(\beta)}{{\rm d}\beta}\,.
\end{equation}
Since the measured value of \(M_{s\bar s}\) has small deviations from
LoCP, as shown in Fig.~\ref{fig:Mss}(a), we corrected it using lowest
order chiral perturbation theory. The corrected mass parameter used in
the analysis and to obtain \(R_m\) is shown in Fig.~\ref{fig:Mss}(b).

\begin{figure}[b]
  \begin{center}
    \begin{tabular}{cc}
      \colorbox{white}{\includegraphics[width=0.38\textwidth]{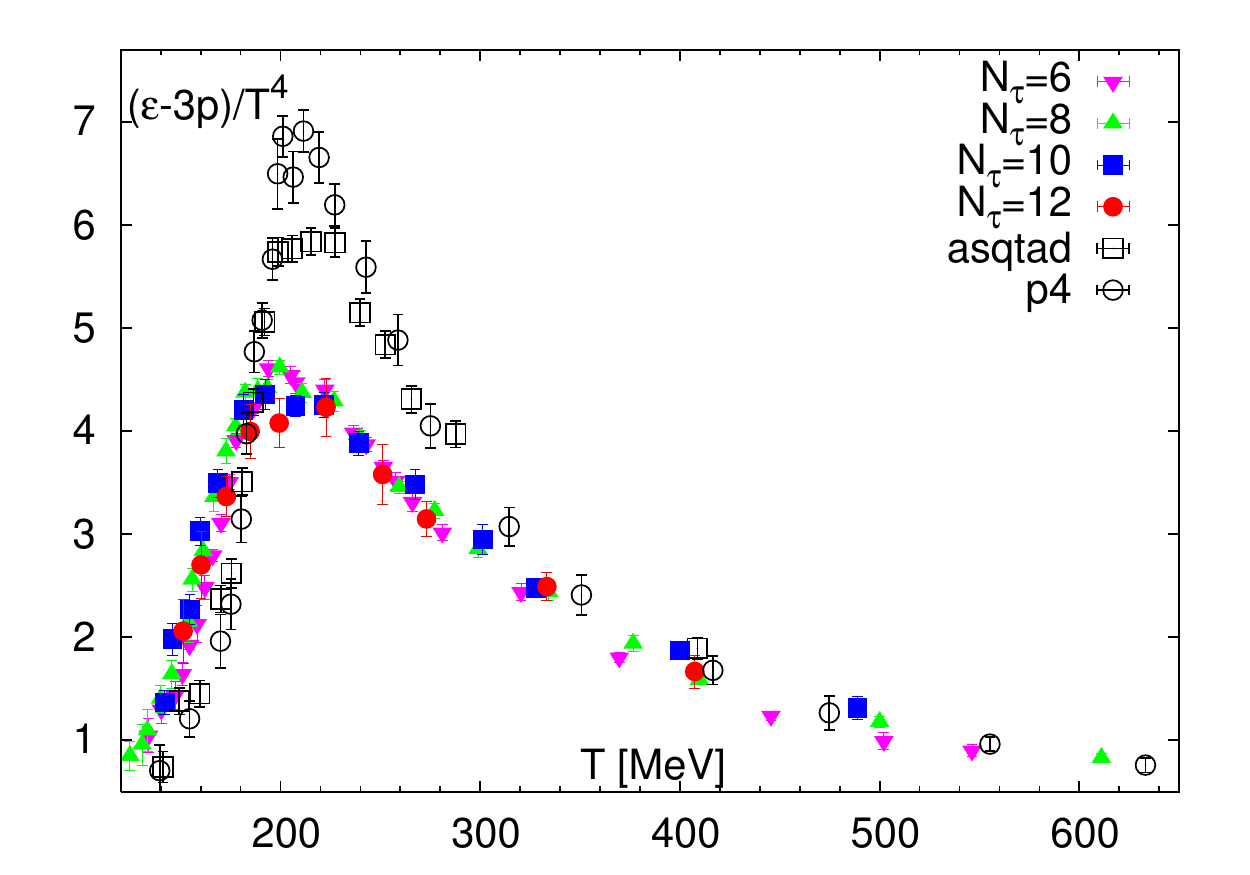}}&
      \colorbox{white}{\includegraphics[width=0.38\textwidth]{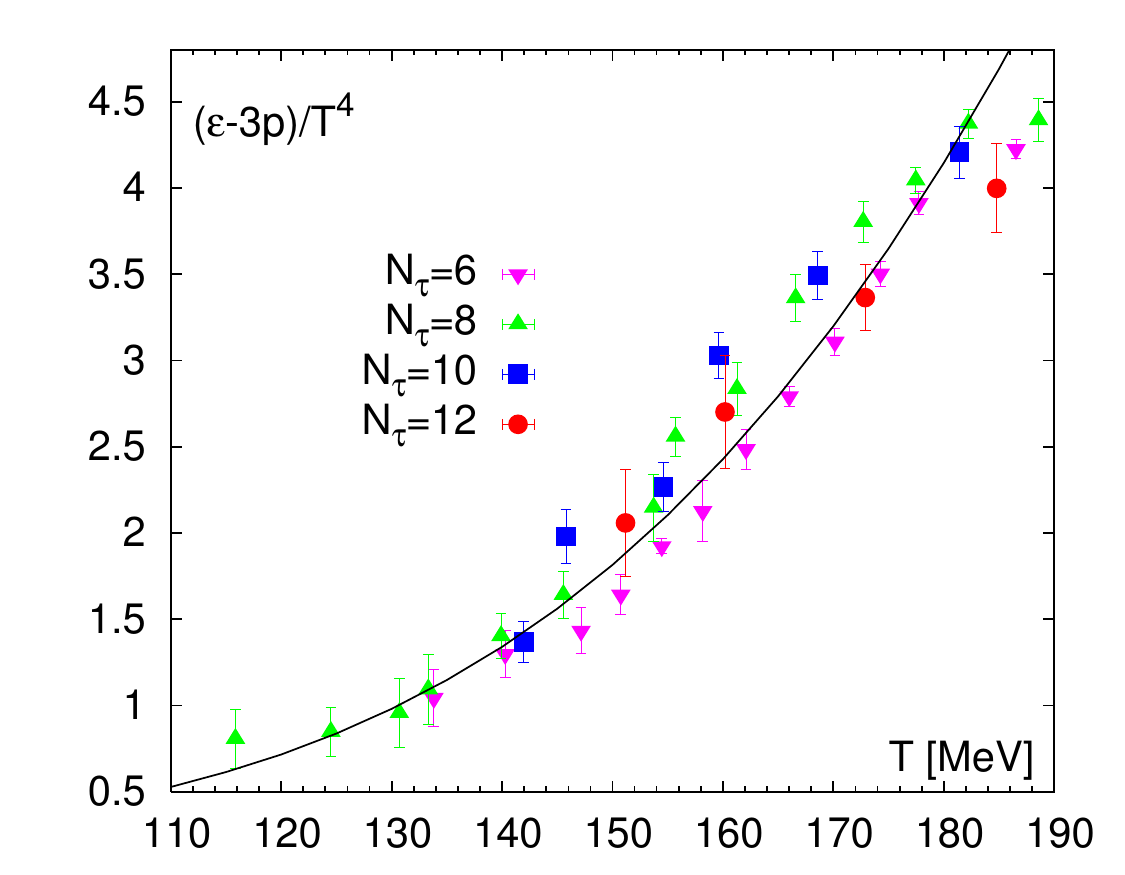}}\\
      (a)&(b)
    \end{tabular}
  \end{center}
  \caption{(a) The trace anomaly calculated at four different
    \(N_\tau\) compared to our previous
    calculations~\cite{Cheng:2007jq,Bazavov:2009zn,Petreczky:2009at}. (b)
    A magnification of the low temperature region with the results
    from a HRG model including all hadrons in the Particle Data
    Book~\cite{Beringer:1900zz} with masses less than
    \(2.5~\text{GeV}\) shown as a solid line.}
  \label{fig:traceanomaly}
\end{figure}

All our data for the trace anomaly are shown in
Fig.~\ref{fig:traceanomaly}(a) and compared to our previous
calculations using the p4 and asqtad lattice actions.  The noticeable
difference is that the peak in the HISQ/tree data is lower and shifted
to the left.  In Fig.~\ref{fig:traceanomaly}(b), we show that the
results for \(T \lesssim 145~\text{MeV}\) at different \(N_\tau\) are
consistent and agree with those obtained using the hadron resonance
gas (HRG) model.

\section{Continuum Extrapolation}

\begin{figure}[t]
  \begin{center}
    \begin{tabular}{cc}
      \colorbox{white}{\includegraphics[width=0.38\textwidth]{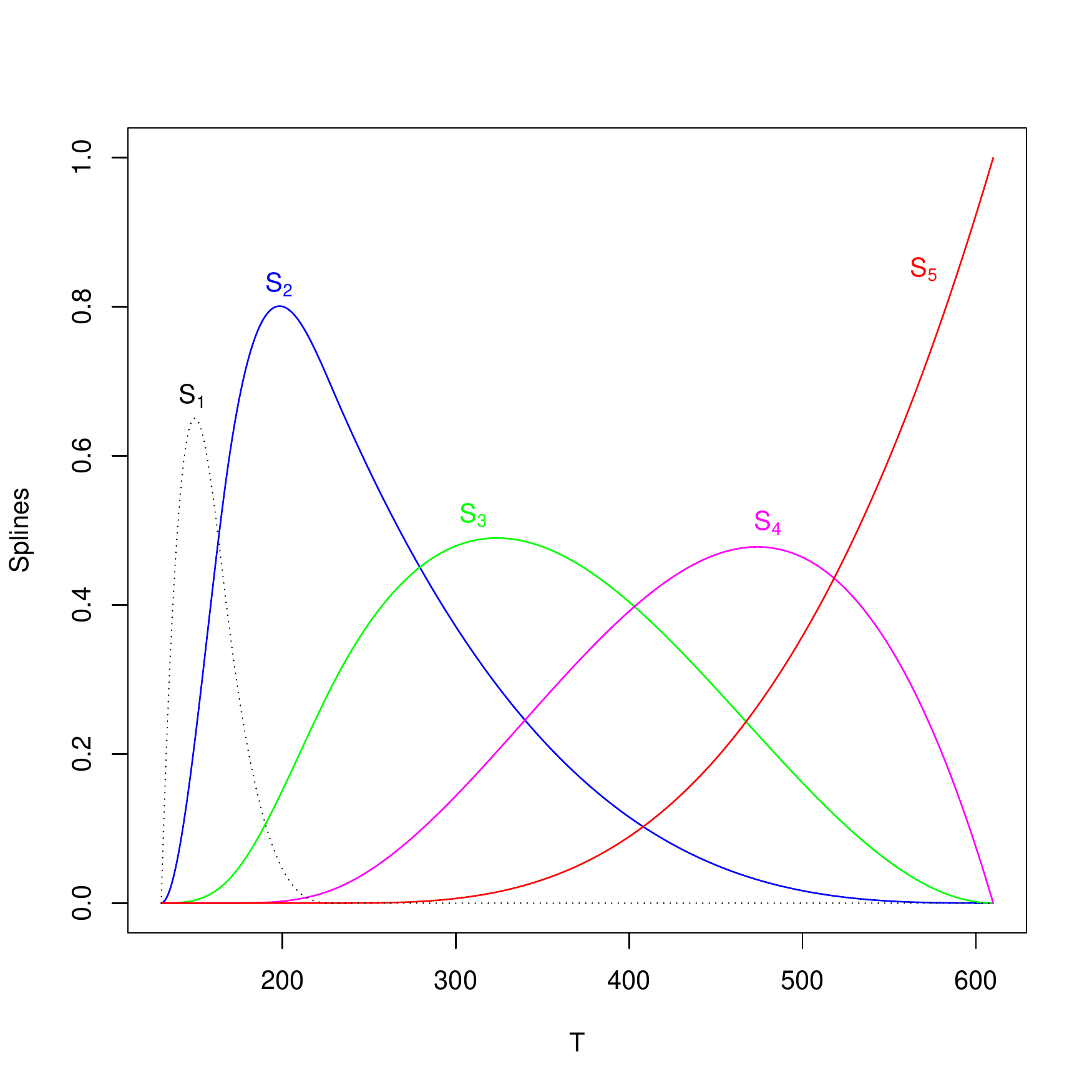}}&
      \colorbox{white}{\includegraphics[width=0.38\textwidth]{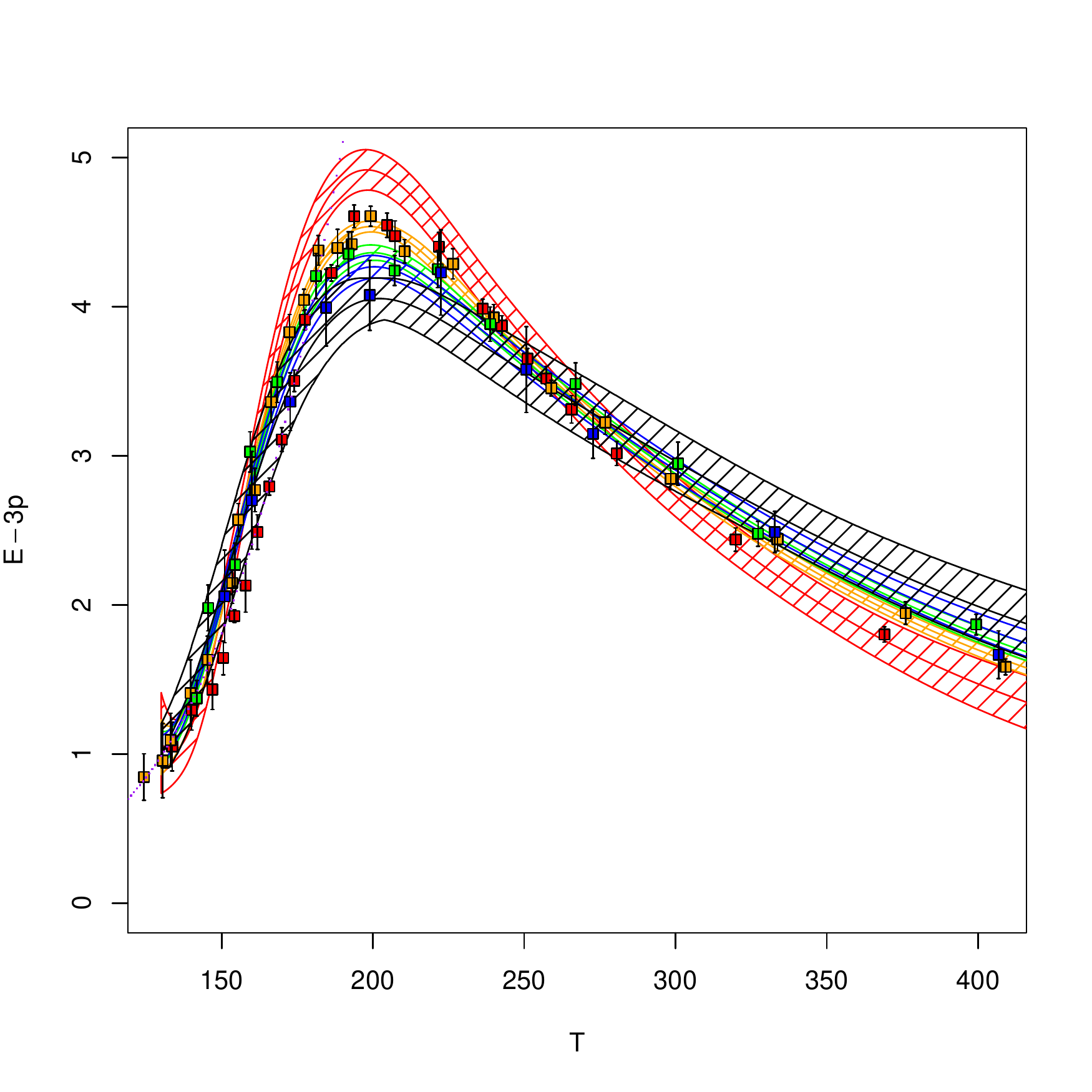}}\\
      (a)&(b)
    \end{tabular}
  \end{center}
  \caption{(a) A B-spline basis with two internal knots that are
    constrained to go to zero at \(130~\rm{MeV}\). (b) The
    extrapolation of the lattice data at \(N_\tau = 8\), \(10\) and
    \(12\) to the continuum is shown as a black band. The fit results
    at the various \(N_\tau\) are shown as colored bands.}
  \label{fig:Bsplines}
\end{figure}

The quality of the lattice data is sufficiently good that various
strategies for extrapolating them to the continuum limit agree within
errors.  The one with least free parameters is a simultaneous fit that
interpolates the data in \(T\) and extrapolates in \(1/N_\tau^2\to0\).
This is implemented by choosing a basis of cubic splines that enforces
the continuity of the function along with its first and second
derivatives in \(T\). An example of a B-spline basis with two knots is
shown in Fig.~\ref{fig:Bsplines}(a).

We explore fits with coefficients of the basis functions chosen as linear or
quadratic functions of \(1/N_\tau^2\), thus enforcing the same
continuity requirements on the extrapolation.  To stabilize our fits,
we match the value and the first temperature-derivative of the
continuum extrapolated value on to the HRG results at
\(T=130~\text{MeV}\).  Since with \(n\) internal knots, the B-spline
basis has \(n+4\) basis elements, the HRG condition at 130 MeV reduces
the number of continuum coefficients to \(n+2\). For the extrapolation
linear in \(1/N_\tau^2\), we have an additional \(n+4\) coefficients
and the choice of \(n\) knot positions, for a total of \(3n+6\)
degrees of freedom (d.o.f).  The fit was performed by minimizing the
uncorrelated \(\chi^2\) and \(n\) was chosen according to the Akaike
Information Criterion (AIC) of minimizing \(\chi^2+2\times\text{d.o.f}\).
All the calculations were done using the statistical program
R~\cite{Rcore} and its Hmisc package~\cite{RpackageHmisc}.

To estimate the errors in the fit parameters, we generated 20,001 synthetic data
sets drawn assuming a normal distribution with variance given by the
estimated error on each measurement. We assigned an independent conservative 10\% 
error on both the value and slope of the HRG at \(T=130~\text{MeV}\).
To the bootstrap error we linearly added a 2\% error to account for scale
uncertainty in determining \(T\). 

An extrapolation ansatz just linear in \(1/N_\tau^2\) does not fit the
\(N_\tau=6\) data well. Our best estimate is obtained using a fit with 2
knots and linear in \(1/N_\tau^2\) to the \(N_\tau = 8\), \(10\) and
\(12\) data at \(T  \lesssim 400~\text{MeV}\).  To stabilize the fit
at the upper end, we included the \(N_\tau=8\) data up to \(T =
610~\text{MeV}\).  This final extrapolation is shown in
Fig.~\ref{fig:Bsplines}(b).

\section{Results and Conclusions}

\begin{figure}[t]
  \begin{center}
    \begin{tabular}{cc}
      \colorbox{white}{\includegraphics[width=0.38\textwidth]{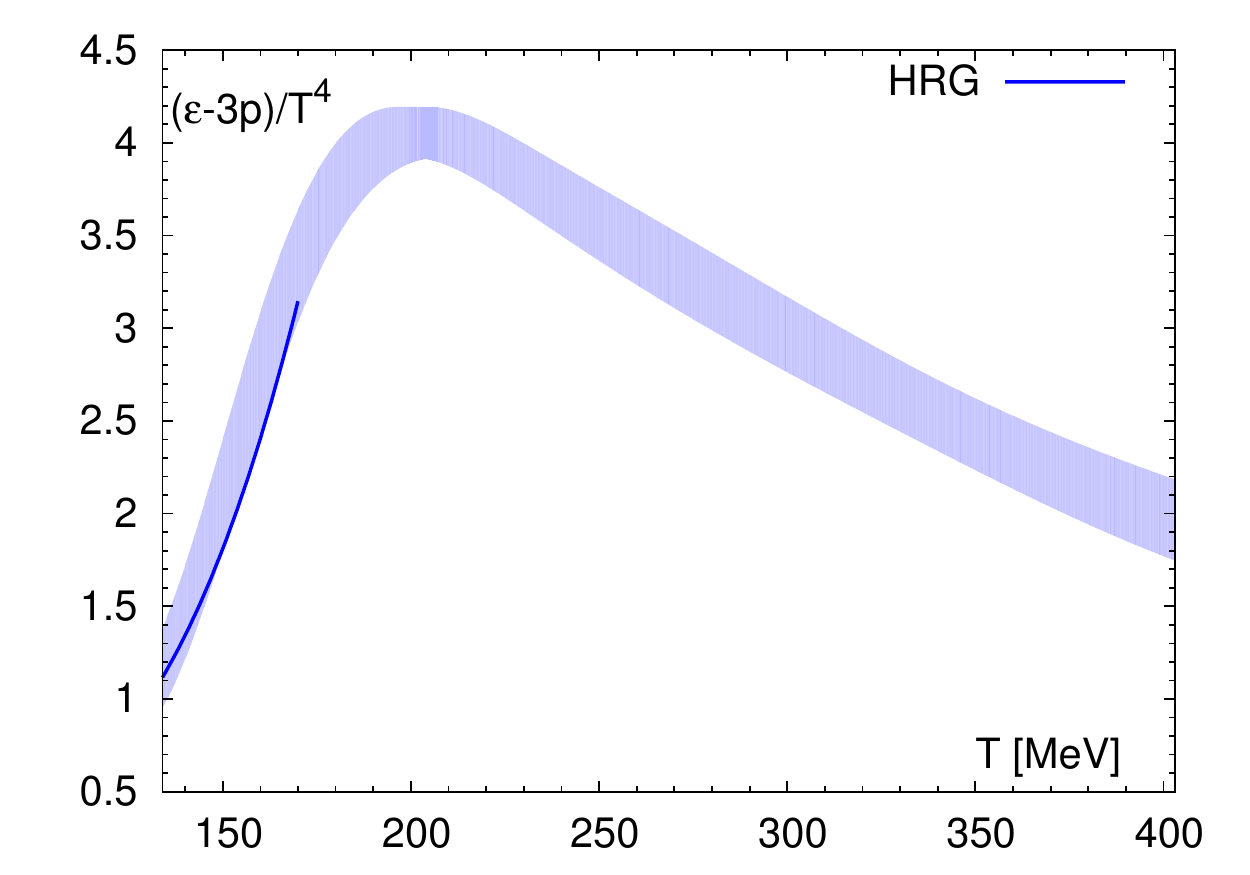}}&
      \colorbox{white}{\includegraphics[width=0.38\textwidth]{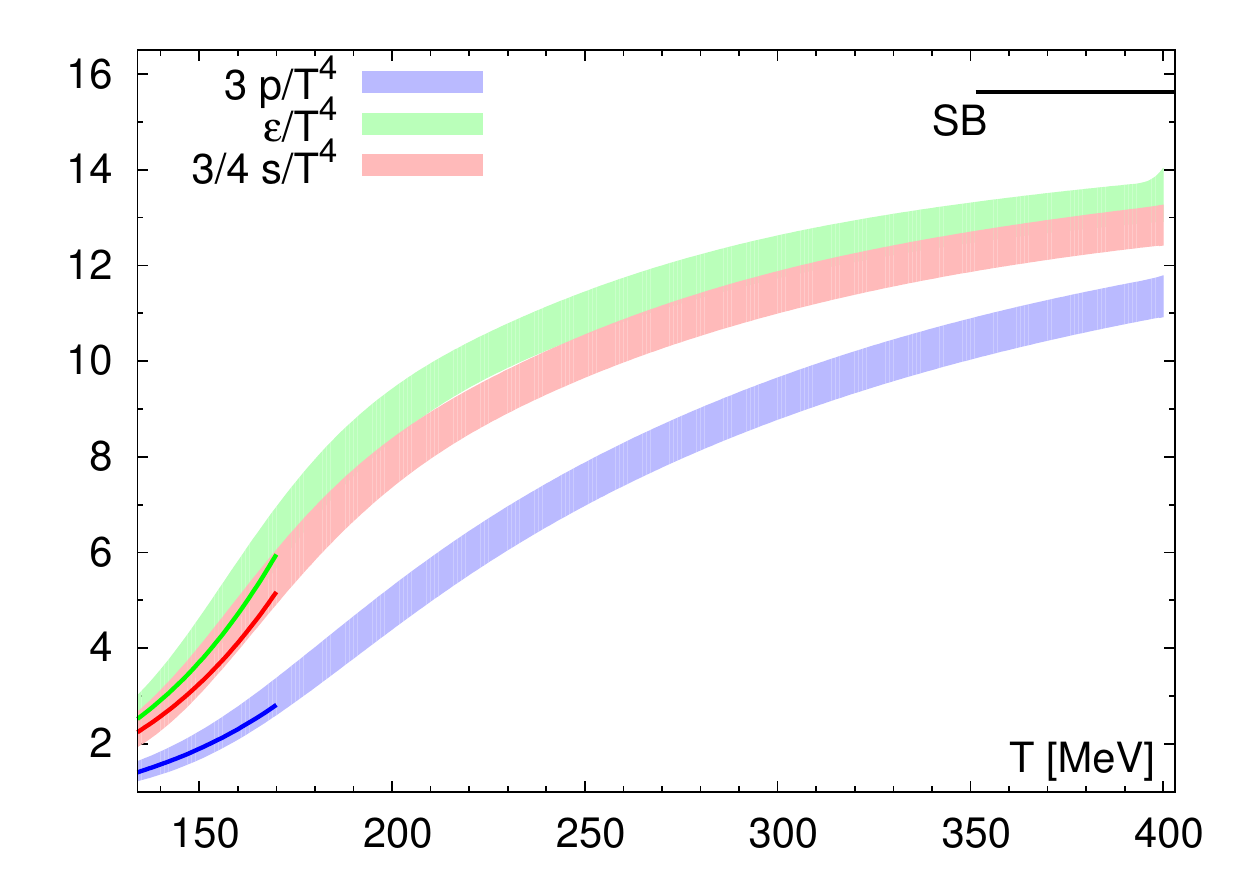}}\\
      (a)&(b)
    \end{tabular}
  \end{center}
  \caption{(a) The continuum extrapolated integration measure (band)
    compared to the HRG estimate (solid line). (b) The continuum
    extrapolated pressure, \(p\), energy density, \(\varepsilon\), and
    entropy density, \(s\), (colored bands) compared to the HRG
    estimate (colored solid lines) and the Stephan-Boltzman value
    (solid black line) for an ideal gas.\looseness-1}
  \label{fig:res}
\end{figure}

\begin{figure}[t]
  \begin{center}
    \begin{tabular}{cc}
      \multicolumn{2}{c}{\colorbox{white}{\includegraphics[width=0.38\textwidth]{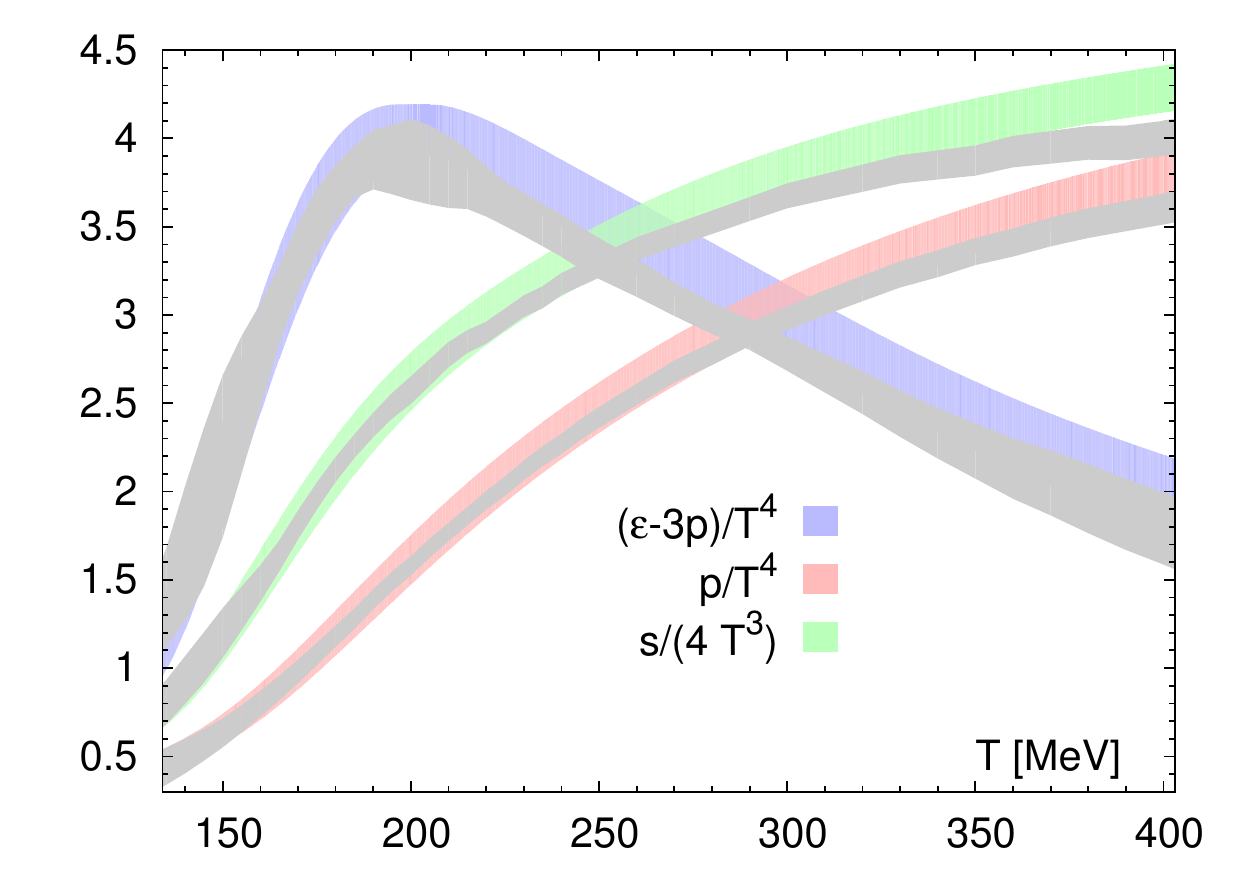}}}\\
      \multicolumn{2}{c}{(a)}\\
      \colorbox{white}{\includegraphics[width=0.38\textwidth]{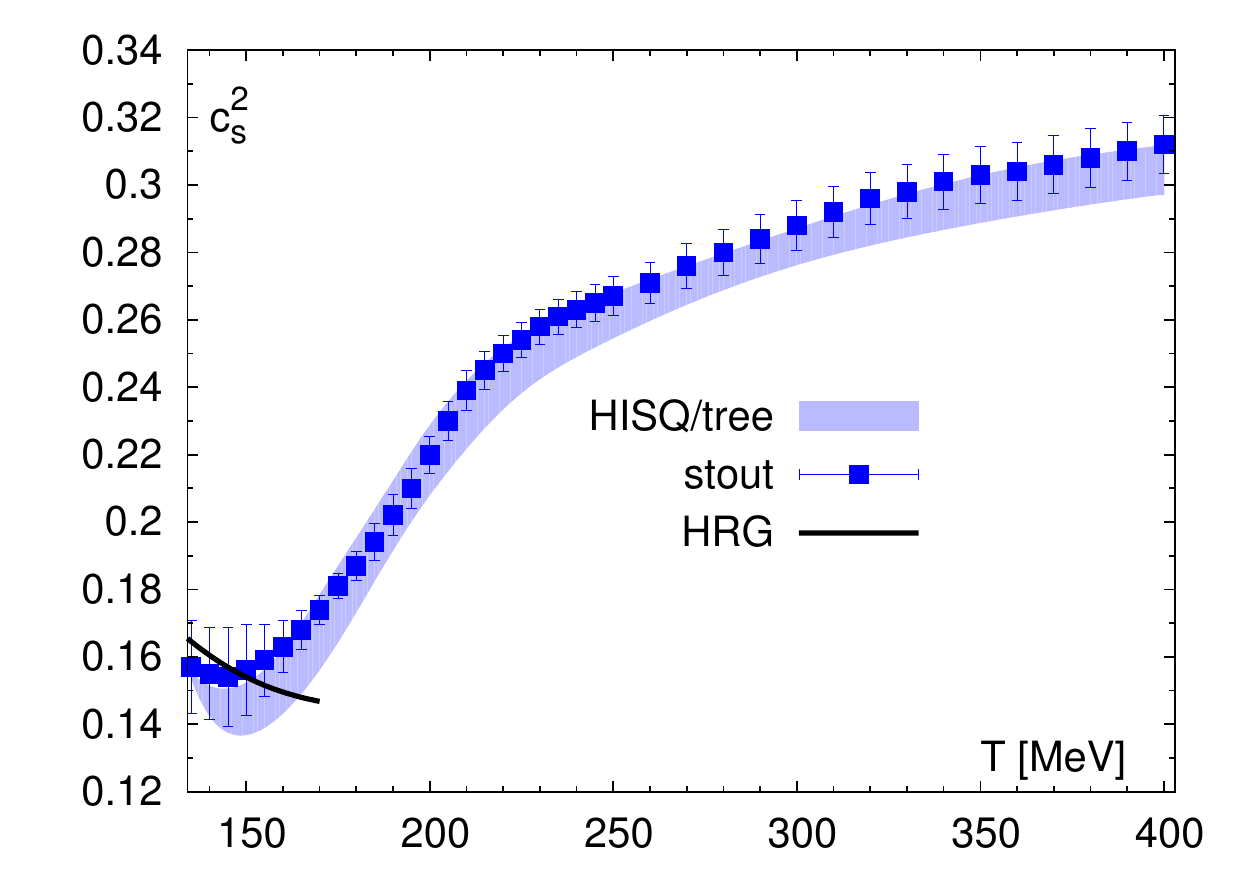}}&
      \colorbox{white}{\includegraphics[width=0.38\textwidth]{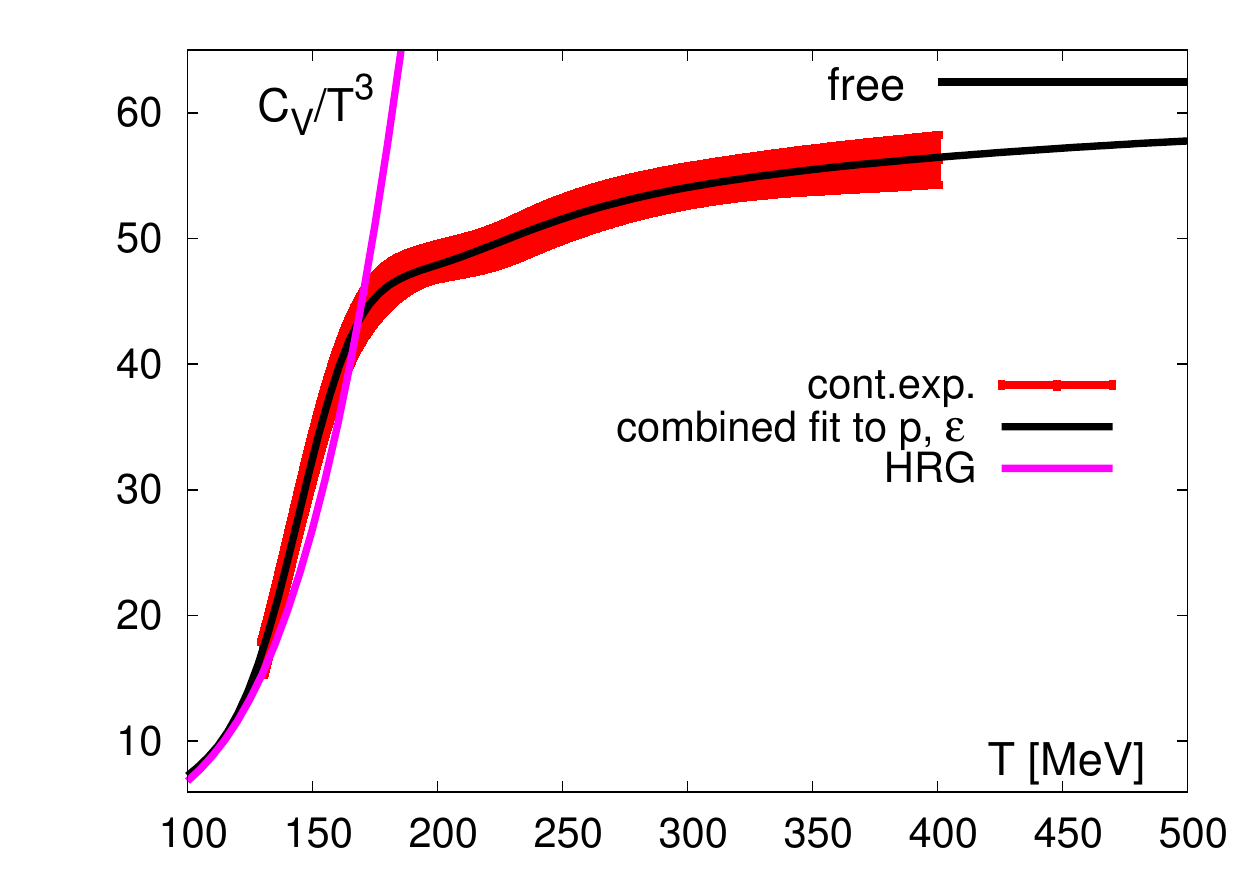}}\\
      (b)&(c)
    \end{tabular}
  \end{center}
  \caption{(a) Comparison of our results for pressure, \(p\), energy
    density, \(\varepsilon\), and entropy density, \(s\), (colored
    bands) with the results from the Wuppertal-Budapest collaboration
    using the stout discretization~\cite{Borsanyi:2013bia} (grey
    bands). (b) Our results for the speed of sound, \(c_s\) (colored
    band) compared to the stout results (data points) and the HRG
    estimate (black solid line). (c) Our results for the specific
    heat, \(C_V\), (colored band) compared to the HRG results (colored
    line) and the value for a non-interacting gas (horizontal black
    line).  Also shown is the result from a parameterization designed
    to the fit the pressure and energy density (black curve).}
  \label{fig:compare}
\end{figure}

As shown in Fig.~\ref{fig:res}, our extrapolated continuum results for
the integration measure, pressure, energy density and the entropy
density agree in value with the HRG estimates for \(T \lesssim
150~\text{MeV}\). The speed of sound and the specific heat shown in
\ref{fig:compare}(b,c), which depend on higher derivatives of the
partition function, are, however, not well predicted by this model.
Figs.~\ref{fig:res}(b) and \ref{fig:compare}(c) also show that the
quark-gluon plasma is still significantly far from a non-interacting
gas even at \(T\approx 400~\text{MeV}\).

In Fig.~\ref{fig:compare}(a) and (b) we comapre our results with those
by the Wuppertal-Budapest collaboration using a stout dicretization of
the fermion action~\cite{Borsanyi:2013bia} and find good agreement.
There may be a small discrepancy for \(T>350~\text{MeV}\) which is
probably unimportant for current heavy ion phenomenology, but needs
further investigation to study the approach to the high temperature
perturbative regime. Lastly, the disagreements with our previous
results for the EOS using the p4 and asqtad actions were due to large
cutoff effects in those calculations.

\bibliographystyle{doiplain}
\bibliography{skeleton}


\end{document}